\documentclass[12pt,preprint]{aastex}
\usepackage{color}

\newcommand{\msun}{{\rm\,M_\odot}}
\newcommand{\Msun}{{\msun}}

\newcommand{\ct}{\citealt}

\begin{document}

\title{Effect of Magnetic Misalignment on Protobinary Evolution} 

\author{Bo Zhao\altaffilmark{1}, Zhi-Yun Li\altaffilmark{1}, Kaitlin M. Kratter\altaffilmark{2,3}}
\altaffiltext{1}{University of Virginia, Astronomy Department, Charlottesville, VA, USA}
\altaffiltext{2}{JILA and CU/NIST, University of Colorado, Boulder, CO, 80309, USA}
\altaffiltext{3}{Hubble Fellow}

\begin{abstract}
The majority of solar-type stars reside in multiple systems, especially
binaries. They form in dense cores of molecular clouds that are
observed to be significantly magnetized. Our previous study shows 
that magnetic braking can tighten the binary separation during 
the protostellar mass accretion phase by removing the angular 
momentum of the accreting material. Recent numerical 
calculations of single star formation have shown that 
misalignment between the magnetic field and rotation axis 
may weaken both magnetic braking and the associated magnetically driven outflows. These two effects allow for disk formation even in strongly magnetized cores. Here we investigate the effects of magnetic field misalignment on the properties of protobinaries. Somewhat surprisingly, the misaligned magnetic field is more efficient at tightening the binary orbit compared to the aligned field. The main reason is that the misalignment weakens the magnetically-driven outflow, which allows more material to accrete onto the binary. Even though the specific angular momentum of this inner material is higher than in the aligned case, it is insufficient to compensate for the additional mass. A corollary of this result is that a weaker field is required to achieve the same degree of inward migration when the field is tilted relative to the rotation axis. Large field misalignment also helps to produce rotationally-supported circumbinary disks even for relatively strong magnetic fields, by weakening the magnetically-dominated structure close to the binary. Our result may provide an explanation for the circumbinary disks detected in recent SMA and ALMA observations.

\end{abstract}

\section{Introduction}
\label{intro}

Stars form in dense cores that are often observed to be significantly  magnetized (\ct{TrolandCrutcher2008}). The spin-up of infalling gas due to angular momentum conservation twists these magnetic field lines, at least in the ideal MHD limit. This twisting and the associated magnetically driven outflow transport angular momentum outward from the forming star. For realistic levels of magnetization, both analytical and numerical work has shown that this magnetic braking (in the ideal MHD limit) suppresses the formation of rotationally supported disks (\ct{Allen+2003}; \ct{Galli+2006}; \ct{PriceBate2007}; \ct{MellonLi2008}; \ct{HennebelleFromang2008}; \ct{DappBasu2010}; \ct{Krasnopolsky+2012}). 
However, Keplerian disks are routinely observed around evolved Class II objects(e.g. \ct{WilliamsCieza2011}), and increasingly around Class I (\ct{Jorgensen+2009}; \ct{Enoch+2009}; \ct{Takakuwa+2012}), perhaps even Class 0 sources (\ct{Tobin+2012}).

The aforementioned studies have all assumed uniform rotation profiles aligned with a uniform magnetic field. Observed cores, however, have both turbulent velocity profiles \citep{Goodman+1993} and likely non-uniform, misaligned fields . Recent CARMA survey of low mass Class 0 protostars indicates that a number of sources have substantial misalignment between the magnetic field and bipolar outflow axis, which is often taken as a proxy for the rotation axis (\ct{Hull+2013}). 
More recent numerical simulations have shown that misalignment between the field and rotation axis allows for the formation of extended disk in moderately magnetized cores due to less efficient magnetic braking and weaker outflows (\ct{Joos+2012}; \ct{Li+2013}). 

In this work we consider the influence of field alignment on the
formation of binary stars. Since both the formation and orbital
parameters of binaries are linked to the angular momentum evolution in
the natal core (e.g., \ct{Hanawa+2010}), we expect  the field geometry to strongly influence the configuration of young binaries. Binary formation may be the dominant channel for star formation. Both field stars and pre-main sequence stars show high binary fractions. For field stars of $\gtrsim 1 M_\odot$, the binary fraction is $\gtrsim 50\%$ \citep{DuquennoyMayor1991,Raghavan+2010,Janson+2012}. Young clusters show multiplicity fraction in excess of $60-70\%$ \citep{ReipurthZinnecker1993,Mathieu+2000,Duchene+2004,Duchene+2007,Kraus+2011}. The high fraction of multiples at birth may indicate that the majority of stars are formed in multiple systems, especially binaries. 

In \citet{ZhaoLi2013} (hereafter ZL13), we studied the influence of
magnetic fields on the growth and orbital evolution of protobinary
seeds embedded in a magnetized core. We found that for magnetic fields
aligned with the core rotation axis, strong magnetic braking
efficiently shrinks the binary separation on a timescale shorter than
$10^{5}$~yrs by removing angular momentum from the infalling gas. In
contrast, similar simulations of binary formation in the hydrodynamic
limit find that the binary separations typically increase after birth
due to the accretion of high angular momentum material
\citep{Kratter+2010}. If most binaries form on wide ($\sim 500-1000$
AU) orbits in magnetized cores, magnetic braking could provide an
explanation for the non-detection of closer systems ($\sim
150-550$~AU) in the Class 0 phase, as claimed by \citet{Maury+2010}
and \citet{Enoch+2011}. These same systems might contribute to the
higher binary fraction for Class I sources  (as high as $\sim 18\%$ in
the same separation range \citep{Connelley+2008}. 

In this paper, we extend our study of protobinaries to include cores with tilted magnetic fields. We compare the accretion history and orbital evolution of misaligned and aligned systems. We use the MHD version of ENZO AMR code (\ct{BryanNorman1997}; \ct{O'Shea+2004}; \ct{WangAbel2009}; \ct{Wang+2010}) to run a series of simulations analogous to ZL13 but with misaligned fields. We show that the change in the efficiency of magnetic braking in the binary case is not entirely analogous to that effect in the single star case.  In \S~\ref{setup}, we discuss the initial setup for the binary seeds and the rotating magnetized gas envelope. Our main results are presented in \S~\ref{result}, where we show that  the field misalignment reduces, rather than increases, the binary separation compared to the case where the field and rotation axes are aligned. We demonstrate how this result is echoed by the changes in disk and outflow morphology in \S~\ref{diskoutflow}. Finally, we discuss the implications of our results for observed systems in \S~\ref{discuss}.

\section{Problem Setup}
\label{setup}

To facilitate comparison with ZL13, we mimic their initial conditions. We begin with a pair of binary seeds at the center of the dense core, and vary both the strength and the orientation of the global magnetic field with respect to the rotation axis. For completeness,  we briefly summarize the initial conditions, and describe the implementation of different magnetic field inclinations. 

We initialize the protobinary envelope with a self-similar density profile (\ct{Shu1977}):
\begin{equation}\label{eq:sis}
\rho (r)= {{A c_s^2} \over {4 \pi G r^2}},
\end{equation}
where $c_s$ is the isothermal sound speed, and $A$ an over-density parameter. Employing self-similar initial conditions provides a powerful check on  numerical solutions. We adopt an over-density parameter $A=4$, corresponding to a ratio of thermal to gravitational energy of $\alpha=3/(2A)=0.375$. The mass enclosed within any radius $r$ is
\begin{equation}\label{eq:mass}
M(r)={{A c_s^2} \over G } r.
\end{equation}
As in ZL13, the total core gas mass is $M_{\rm tot}=1.2\Msun$, with a core radius $R=10^{17}$~cm and isothermal sound speed $c_s=0.2$~km/s (corresponding to a temperature of $\sim 10$~K). The rotation speed of the core is chosen as $v_{\phi}=v_0$~sin~$\theta$ (where $\theta$ is the polar angle measured from the rotation axis, and $v_0=c_s$), which preserves the self-similar collapse. Such a rotation profile corresponds to a ratio of rotational and gravitational energy $\beta=(v_0/c_s)^2/(3A) \approx 0.083$. Note that this is somewhat higher than used in other works (e.g. \ct{Machida+2010}), but is still within the range inferred by \ct{Goodman+1993} from $NH_3$ observations of dense cores.

Unlike ZL13, we now allow the magnetic field to tilt by $\theta_0 = 45^{\circ}$ and $90^{\circ}$ relative to the axis of rotation. The strength of the uni-directional magnetic field has the same initial profile of
\begin{equation}
B_z(\varpi)={ {A c_s^2} \over {\sqrt{G} \lambda} } {1 \over {\varpi}+r_h},
\end{equation}
where the $\lambda$ is the dimensionless mass-to-flux ratio of the envelope in units of the critical value for a magnetically supported core $(2{\pi}G^{1/2})^{-1}$, $r_h$ is the softening parameter to avoid the singularity at the origin. Note that the field strength decreases away from the magnetic axis as $1/\varpi$ (where $\varpi$ is the cylindrical radius relative to the magnetic axis), so that the mass-to-flux ratio is constant spatially. We perform 10 simulations: we have two tilt angles $45^{\circ}$ and $90^{\circ}$ for 5 levels of initial magnetization of $\lambda=2$, $4$, $8$, $ 16$, and $32$. These are compared with the aligned cases (with $\theta_0=0^{\circ}$) of ZL13.

The binary stars are modeled in the same way as in ZL13 using sink particles. The simulations are initialized with two sink particles separated by $a \approx 246$~AU at the center of the protostellar envelope (see Eq. 5 of ZL13). The seeds have a small initial mass of $0.05 \Msun$ each and thus do not significantly modify the core potential. The sinks are allowed to accrete mass from the surroundings based on the modified Bondi-Hoyle formula (see \ct{Ruffert1994}).

\section{Results}
\label{result}

\subsection{Protobinary Migration}
\label{separation}

\begin{figure}
\epsscale{0.9}
\plotone{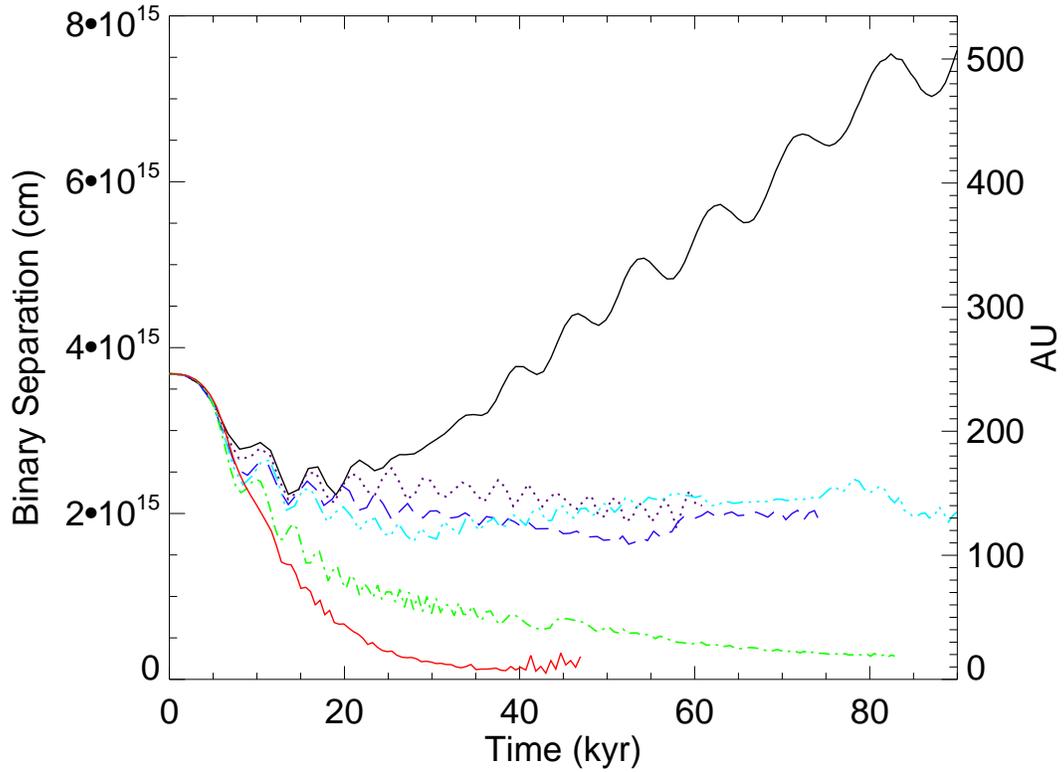}
\caption{Evolution of binary separation with time for HD (black solid), $\lambda=32$ (purple dotted), $\lambda=16$ (blue long-dashed), $\lambda=8$ (teal dash-dot-dot-dotted), $\lambda=4$ (green dash-dotted), and $\lambda=2$ (red thick solid) cases. All magnetic cases have tilt angle $\theta_0=90^{\circ}$.}
\label{inclbinsep}
\end{figure}

We first investigate the effect of field misalignment on the evolution of the binary orbit. Varying the magnetic field strength affects
the binary separation more than varying the alignment. We show the
influence of field strength on separation for the orthogonal case
($\theta_0=90^{\circ}$) in Fig.~\ref{inclbinsep}. As in ZL13, the
binary separation  decreases with increasing magnetic field
strength. Contrary to the expectation based on the single star 
case, weaker braking does not inhibit the shrinking of the binary orbit.

\begin{figure}
\epsscale{0.9}
\plotone{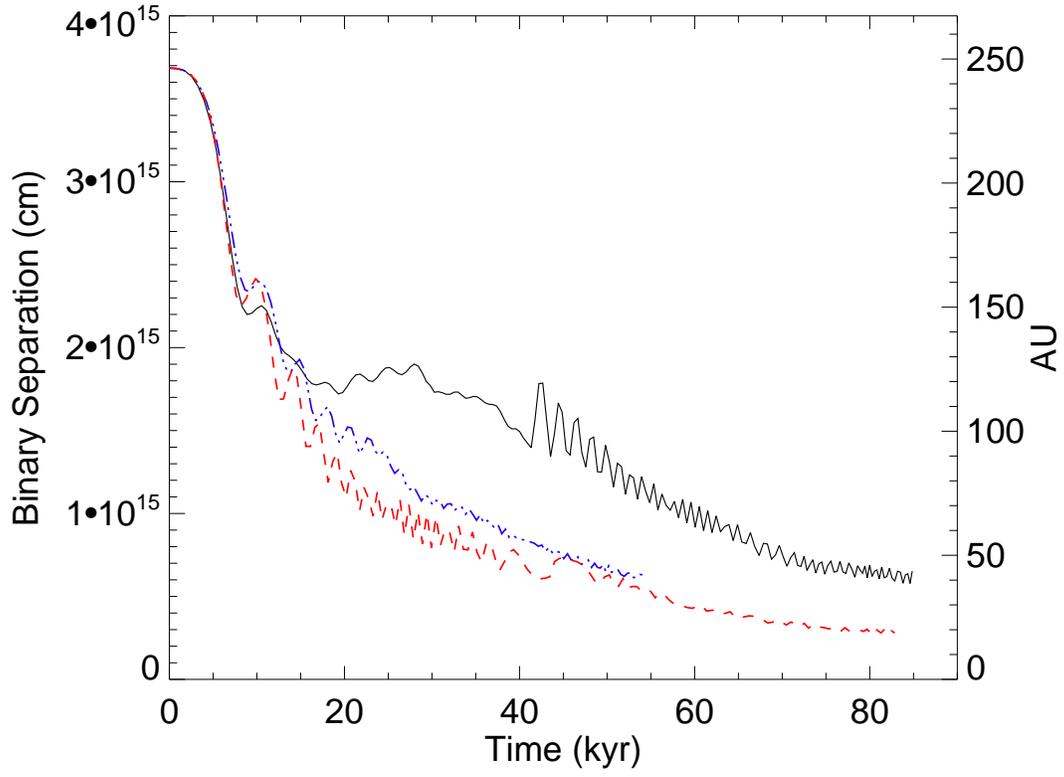}
\caption{Evolution of binary separation with time for $\lambda=4$ cases with different tilt angles: $0^{\circ}$ (black solid),  $45^{\circ}$(blue dash-dotted), and $90^{\circ}$ (red dashed).}
\label{lmd4binsep}
\end{figure}

\begin{figure}
\epsscale{0.9}
\plotone{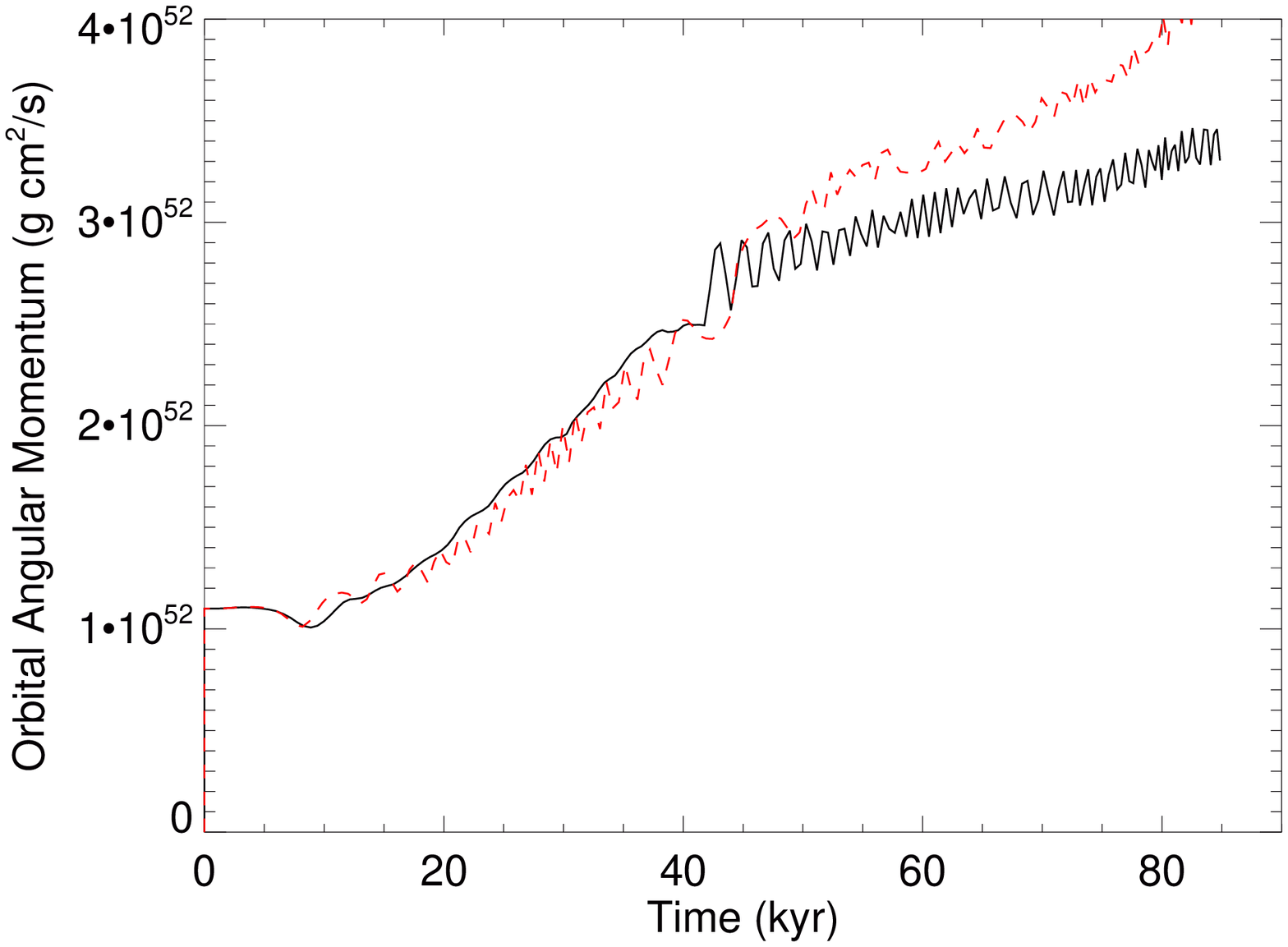}
\caption{Evolution of binary orbital angular momentum with time for
  $\lambda=4$ cases with different inclination angles: $0^{\circ}$
  (black solid) and $90^{\circ}$ (red dashed).}
\label{lmd4am}
\end{figure}

For the same level of magnetization, we find that 
the binary separation is smaller for larger tilt angles at any give time. This is illustrated in Fig.~\ref{lmd4binsep} for $\lambda=4$\footnote{We mainly focus on the two extreme cases with $\theta_0=0^{\circ}$ and $90^{\circ}$. The intermediate case $\theta_0=45^{\circ}$ largely lies in between.}. The difference between the orthogonal case and the aligned case is modest.
One might naively assume that the tighter binary separation in the tilted case is due to stronger magnetic braking. 
However, this would contradict the results of \citet{Joos+2012} and \citet{Li+2013}. To ascertain the cause of angular momentum removal from the binary we plot the evolution of orbital angular momentum for different tilt angles in Fig.~\ref{lmd4am}.

Fig.~\ref{lmd4am} shows that the orbital angular momentum is larger in the orthogonal case than in the aligned case, consistent with the result of \citet{Joos+2012} and \citet{Li+2013} for single star formation. However, the orbit is determined by the angular momentum per unit mass, therefore we need to account for the influence of the field misalignment on stellar accretion.

\begin{figure}
\epsscale{0.9}
\plotone{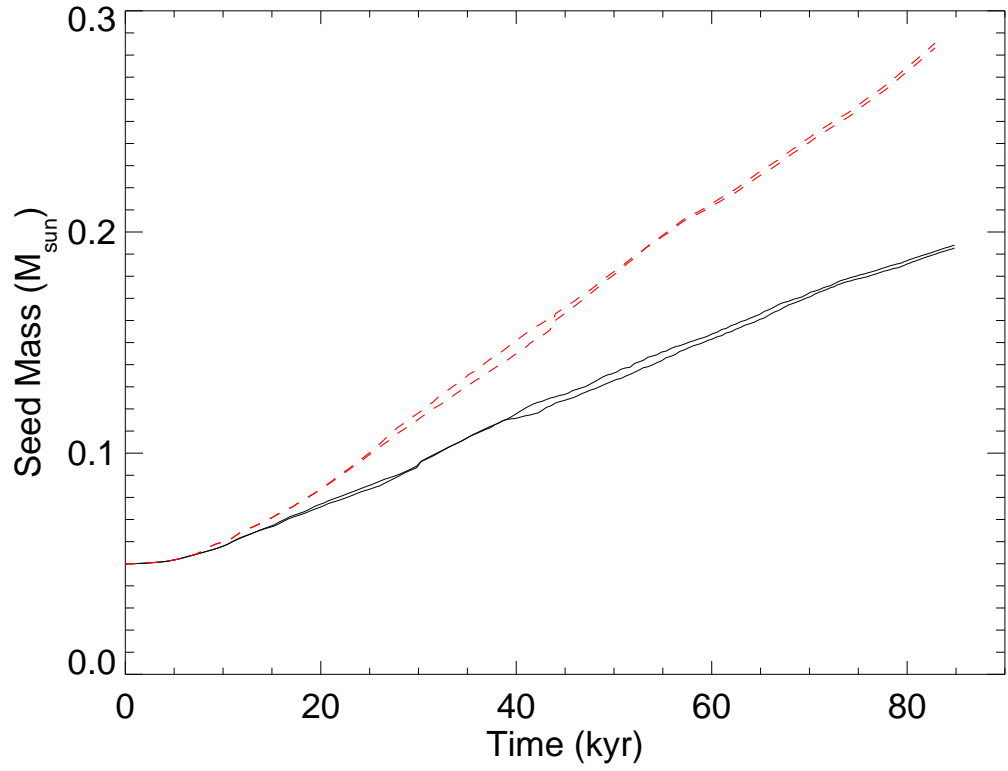}
\caption{Stellar mass (in solar mass) growth with time for $\lambda=4$ cases with different inclination angles of magnetic field: $0^{\circ}$ (black solid) and $90^{\circ}$ (red dashed). In all cases, each star is plotted separately.}
\label{lmd4mass}
\end{figure}

We plot in Fig.~\ref{lmd4mass} the stellar mass as a function of time for the $\lambda=4$ case. It is clear that the binaries accrete faster in cases with a larger tilt angle. The difference is less than a factor of $2$ for the two extremes ($\theta=0^{\circ}$ and $90^{\circ}$), but is sufficient to explain the smaller binary separation for the $\theta=90^{\circ}$ case compared to $\theta=0^{\circ}$ case shown in Fig.~\ref{lmd4binsep}. 

For a binary system on a circular orbit, the orbital angular momentum is given by
\begin{equation}
L_{b}={q \over {(1+q)^2}} G^{1/2} {M_b}^{3/2} {a_b}^{1/2},
\end{equation}
where $L_b$, $M_b$, and $a_b$ are the orbital angular momentum, total mass and separation of the binary, respectively, and $q$ is the mass ratio of the two stars. The binary separation scales as $a_b \varpropto {L_b}^2 {M_b}^{-3}$. 
Thus an increase of binary mass by a factor of $1.5$  yields a binary
separation $\sim 3$ times smaller, given the same amount of available
angular momentum. The reduction in orbital separation due to faster
mass accretion is more than enough to offset the orbit widening due 
to the slightly larger orbital angular momentum in the orthogonal case 
(see Fig.~\ref{lmd4am}). The change in accretion rate accounts for the factor of $\sim2$ difference in the binary separation shown in Fig.~\ref{lmd4binsep} between the aligned and orthogonal case. It is largely consistent with the results of \citet{Joos+2012} and \citet{Li+2013}. We now explore how field misalignment affects mass accretion onto the stars.

\subsection{Mass Accretion}
\label{accretion}

\begin{figure}
\epsscale{1.1}
\plottwo{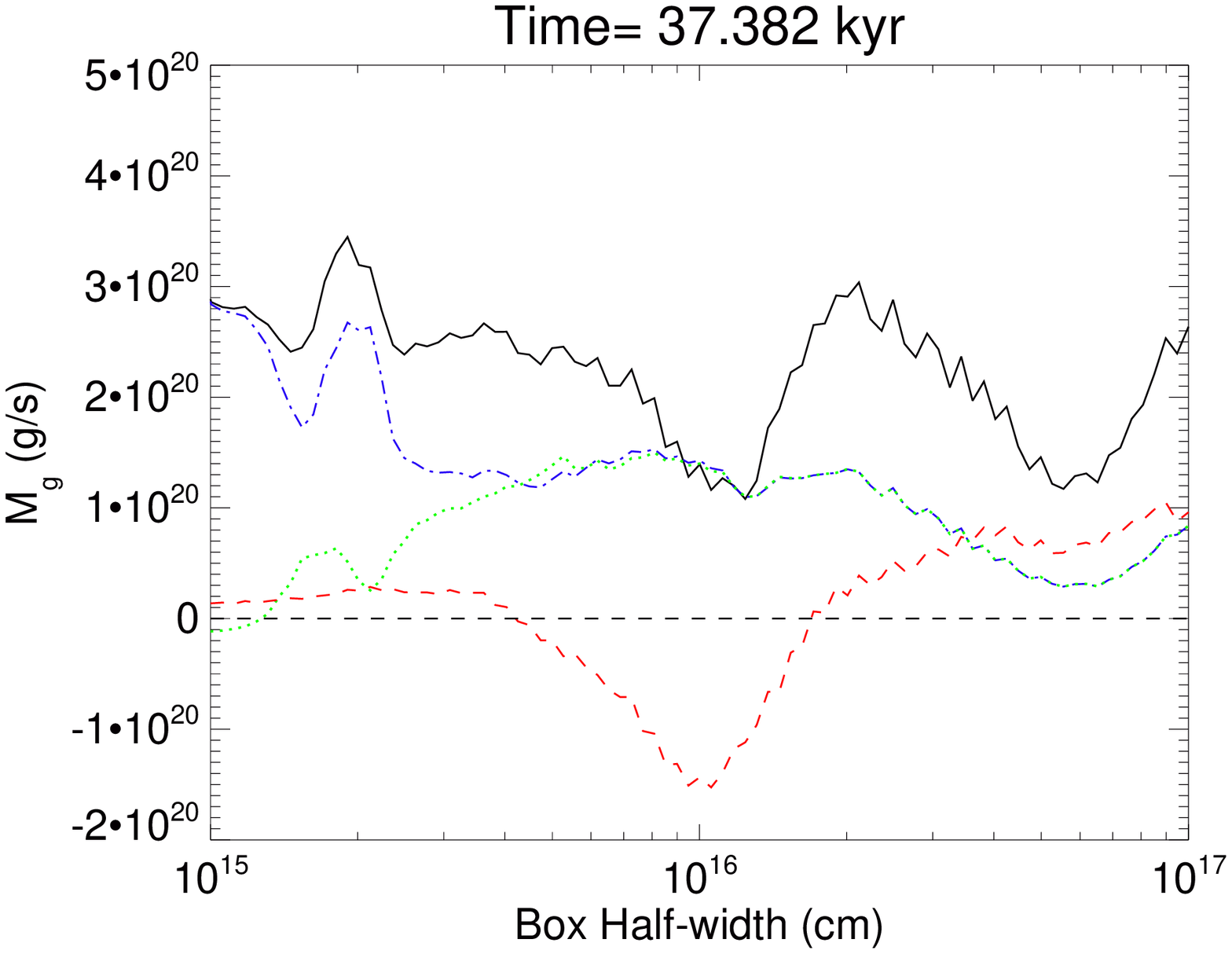}{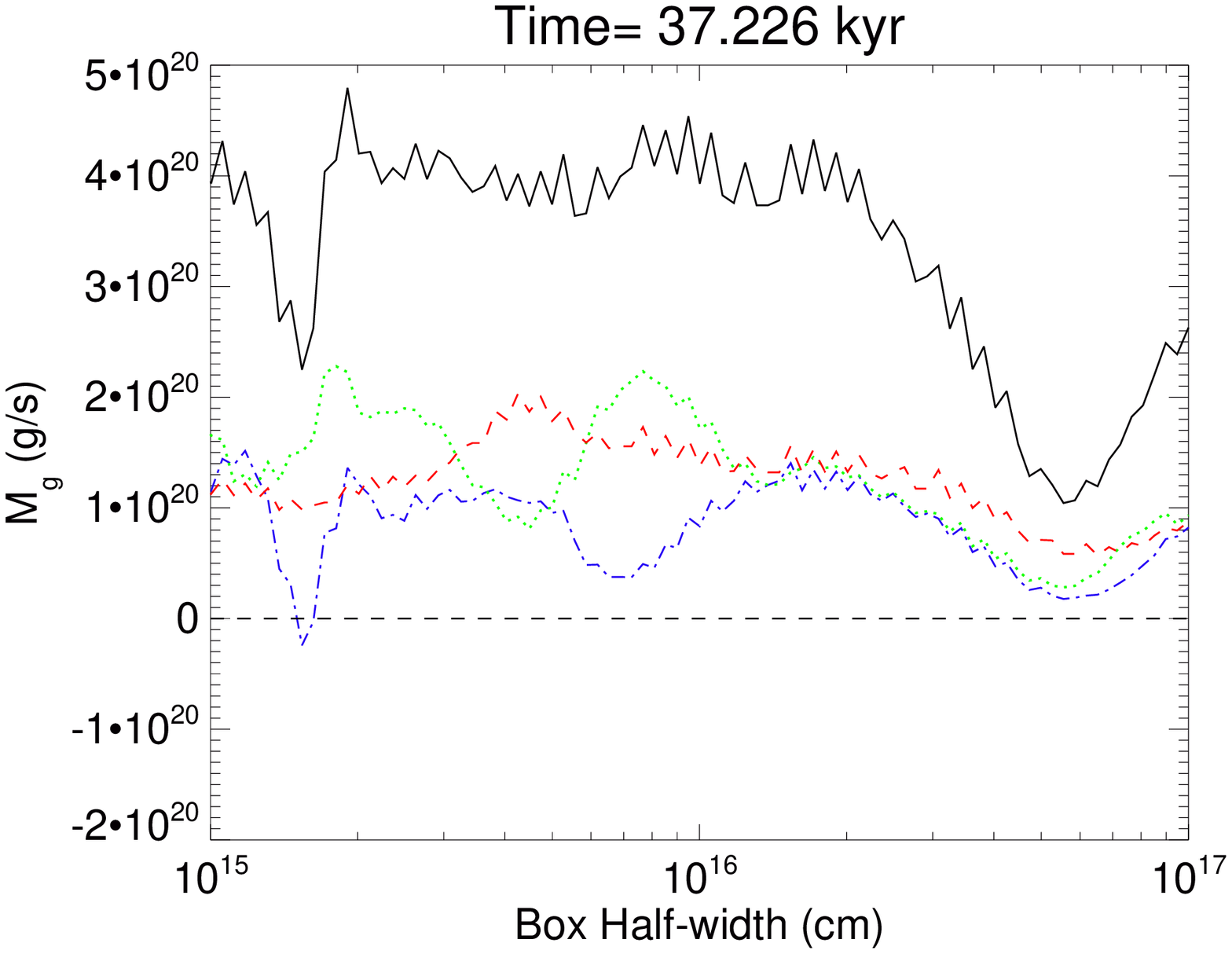}
\caption{The total mass flux (black solid) and its three Cartesian
  components (blue dash-dotted: x-component; green dotted:
  y-component; red dashed: z-component) for cubic boxes of different
  half-width for the $\lambda=4$, at a representative time $t\approx
  37$~kyr. The left panel is the aligned case ($0^{\circ}$) and the
  right panel is the orthogonal case ($90^{\circ}$). A positive flux
  increases the mass within a volume (inflow) whereas a negative one
  decreases it (outflow).}
\label{figmsfl}
\end{figure}

We have shown that the variation in the mass accretion rate for different field geometries is key in understanding the orbital evolution. The accretion of more low specific angular momentum material in misaligned cases compensates for the weaker magnetic braking. We now investigate the difference in binary mass growth for different tilt angles by measuring the gas flow through different surfaces around the binary stars. Let us consider the mass flux through the surface $S$ of a finite volume $V$, which can be expressed as,
\begin{equation}
\dot{M_g}= \int \rho {\bf v} \cdot d{\bf S}
\end{equation}
Here positive mass flux represents inflow and negative represents outflow, e.g. fluid being advected inward or outward through the specified surface. 
As an example, we show in Fig.~\ref{figmsfl} the distribution of the
mass flux $\dot{M_g}$ and its three Cartesian components for cubic
boxes of different sizes that are centered at the origin\footnote{We
  use cubes rather than spheres for simplicity as these align with the
  AMR grid geometry, and thus require no interpolation between
  refinement levels.}.  As expected, the overall mass infall in the
$90^{\circ}$ case is almost $1.5$ times as large as that in the
$0^{\circ}$ case in the region with box half-width between $2\times
10^{15}$~cm and $2\times 10^{16}$~cm ($\sim 1000$~AU). The main driver
for such difference is the z-component of the mass flux $\dot{M_g}$,
whose value is almost the opposite in the two cases. In the aligned
case, the magnetically-driven outflow dominates the gas dynamics
within a distance of $\sim 4\times 10^{15}$~cm to $\sim 2\times
10^{16}$~cm from the center of mass. This outflow accounts for the
bulk of the overall difference in total mass flux. In the orthogonal
case where there is little toroidal field generated, and thus little
or no outflow, the z-component of the mass flux is mostly positive. 
Both the x and y component of $\dot{M_{\rm g}}$ have similar sign and magnitude in the two cases, except for the innermost region close to the binary stars where the flows are becoming unstable (see section \S~\ref{diskoutflow}).

The larger mass inflow may also be correlated with the topology of the magnetic field. In the aligned field case, the vertical toroidal field is severely pinched along the equator as gas collapses towards the central objects. The resulting magnetic tension force impedes the gas infall, so that one would expect a lower accretion rate as time proceeds. On the contrary, in the orthogonal case, the magnetic tension force has limited influence on the gas accretion. Although the rotating gas winds the magnetic field around the binary seeds, the unwinding reaction by magnetic tension does not stop the gas from flowing along the field lines, which lead directly to the center of the system. Hence, the larger magnetic field inclination boosts the gas inflow around the binary.

\subsection{Angular Momentum Transport}
\label{AMtransport}

Mass flow in and around the binary not only delivers angular momentum directly, but also induces hydrodynamical torques (distinct from magnetic torque), which transport angular momentum. We use a similar approach as in \S~\ref{accretion} to quantify the contributions from different torques to the angular momentum transport. For a finite volume $V$ with surface $S$, the total magnetic torque relative to the origin (from which a radius vector ${\bf r}$ is defined) is 
\begin{equation}
{\bf N}_{m}={1 \over {4\pi}} \int[{\bf r} \times ((\nabla \times {\bf B}) \times {\bf B})]\,dV,
\end{equation}
where the integration is over the volume $V$. Typically, the magnetic torque comes mainly from  magnetic tension rather than the magnetic pressure. The dominant magnetic tension term can be simplified to a surface integral (\ct{MatsumotoTomisaka2004})
\begin{equation}
{\bf N}_t={1 \over {4\pi}} \int ({\bf r} \times {\bf B})({\bf B} \cdot d{\bf S}),
\end{equation}
over the surface $S$ of the volume. This volume-integrated magnetic
torque is to be compared with the rate at which angular momentum is advected
into and out of the volume through fluid motion, 
\begin{equation}
{\bf N}_a=-\int \rho({\bf r} \times {\bf v})({\bf v} \cdot d{\bf S}),
\end{equation}
which will be referred to as the advective torque below. 

Since the initial angular momentum of the protobinary envelope is 
along the $z$-axis,
we will be mainly concerned with the $z-$component of the magnetic
and advective torque, 
\begin{equation}
N_{t,z} =  {1 \over {4\pi}} \int (x B_y - yB_x)({\bf B} \cdot d{\bf S}), 
\end{equation}
and 
\begin{equation}
N_{a,z} = -\int \rho (x v_y - y v_x)({\bf v} \cdot d{\bf S}).
\end{equation} 
The advective torque can be separated into two components $N^{out}_{a,z}$ and $N^{in}_{a,z}$, for flow going out of and into the box, respectively. 

\begin{figure}
\epsscale{1.1}
\plottwo{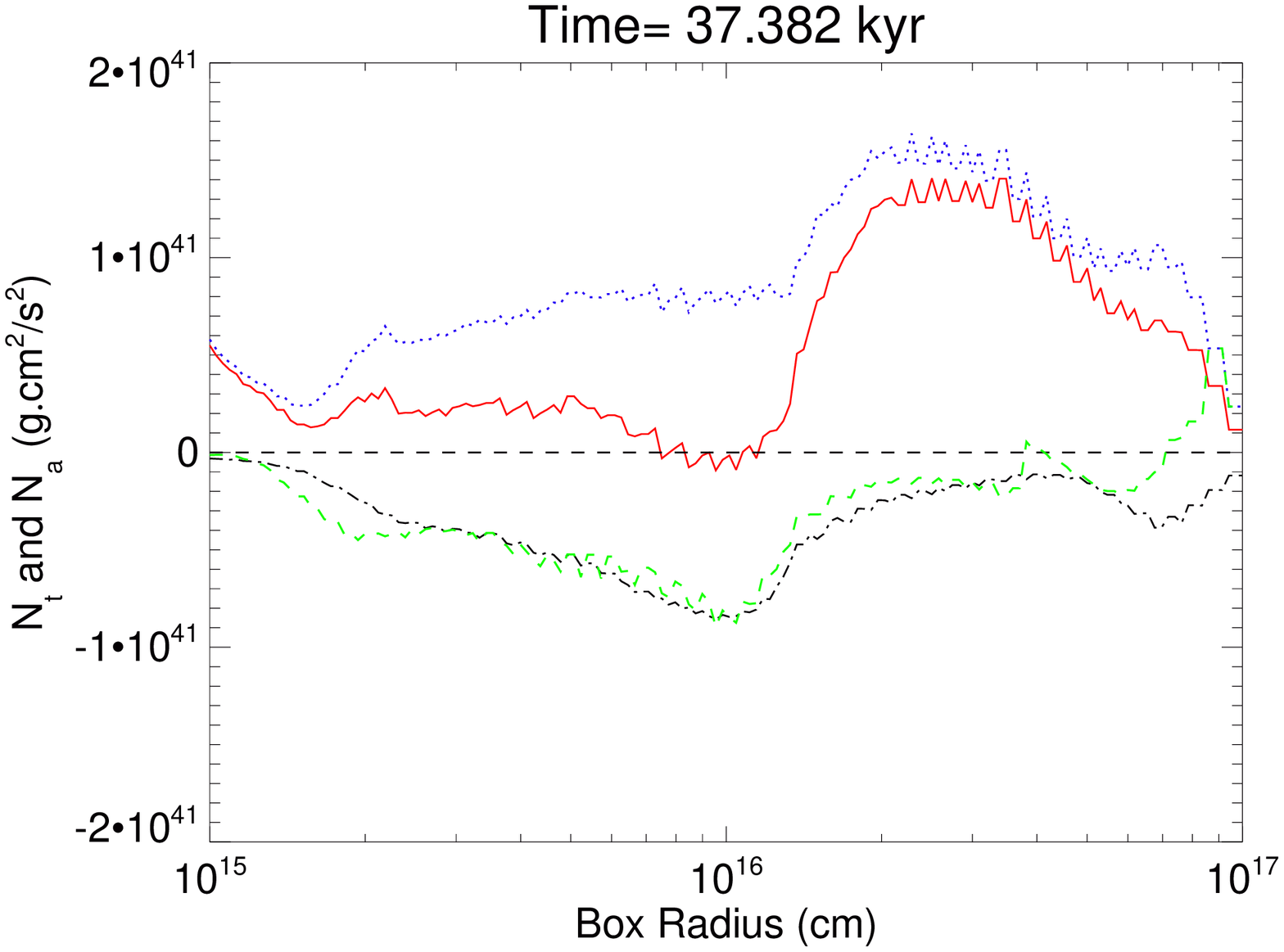}{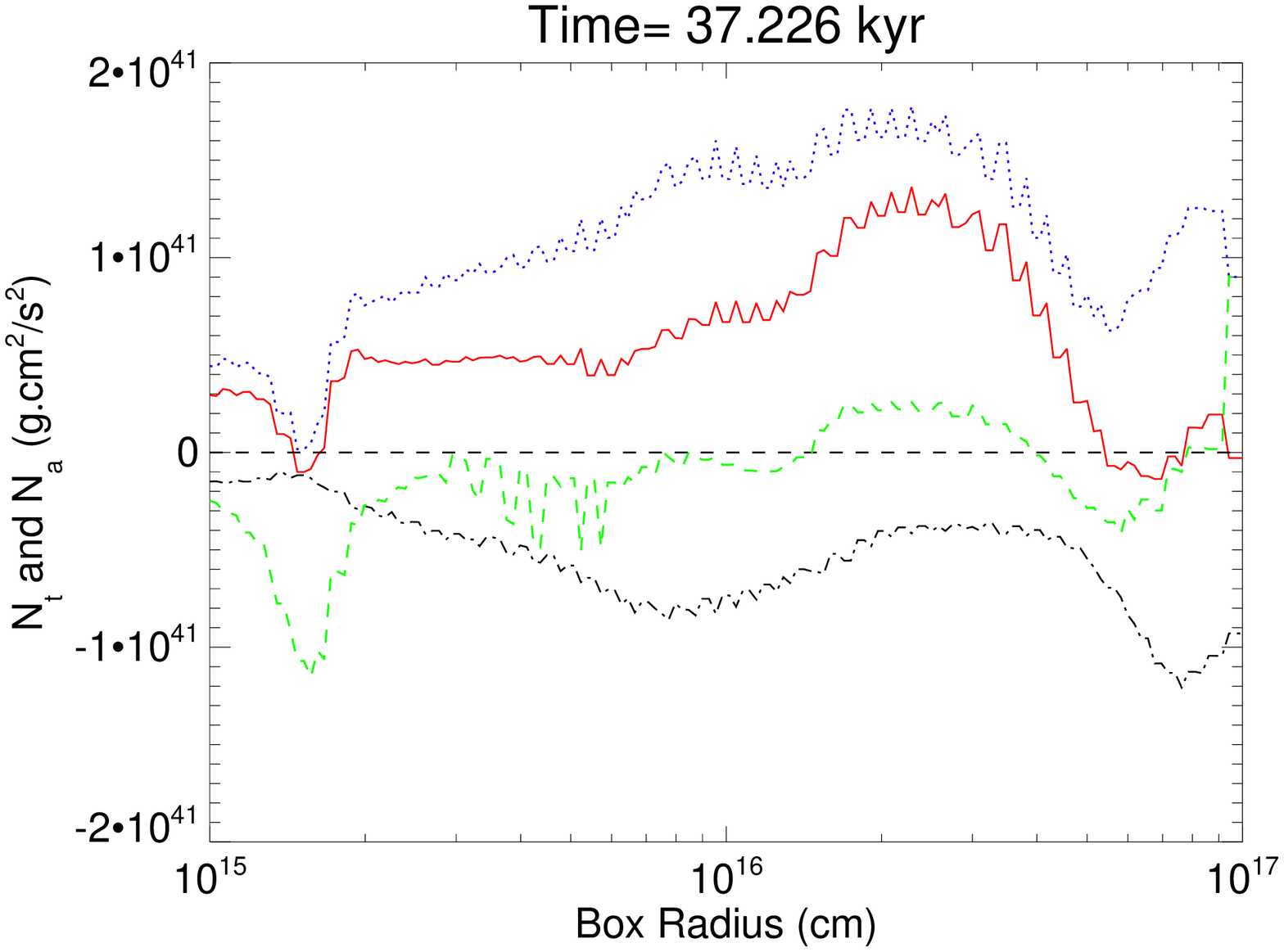}
\caption{The magnetic (black dash-dotted) and advective (blue dotted) torque and the sum of the two (red) for cubic boxes of different half-width $b$ for 
the $\lambda=4$ case, at a representative time $t\approx 37$~kyr. The outflow component of the advective torque is shown in green dashed curve. Left panel is the aligned case ($0^{\circ}$) and the right panel is the orthogonal case ($90^{\circ}$). A positive torque increases the angular momentum within a volume whereas a negative one decreases it. }
\label{figtorq}
\end{figure}

In Fig.~\ref{figtorq} we show the distributions of magnetic and advective torques $N_{t,z}$, $N_{a,z}$, and $N^{out}_{a_z}$ for cubic boxes 
of different sizes that are centered at the origin, again at the time
$t\approx 37$~kyr for both the $0^{\circ}$ and $90^{\circ}$ cases with
the same initial $\lambda=4$. The main difference is the strength of
outflows, which are stronger in the aligned case than  in the
orthogonal case. This morphological difference is obvious from the
column density map and velocity field shown in Fig.~\ref{columnplt} in
\S~\ref{diskoutflow} below. 

The absence of prominent magnetically-driven outflows in the orthogonal
 case helps to maintain a bulk of mass along with their angular momentum to reside near the binaries, lifting the positive change of the angular momentum in the region between $2 \times 10^{15}$~cm and $1\times 10^{16}$~cm. Recall that the similar region also yields a larger mass inflow in the $90^{\circ}$ case. 

\begin{figure}
\epsscale{0.9}
\plotone{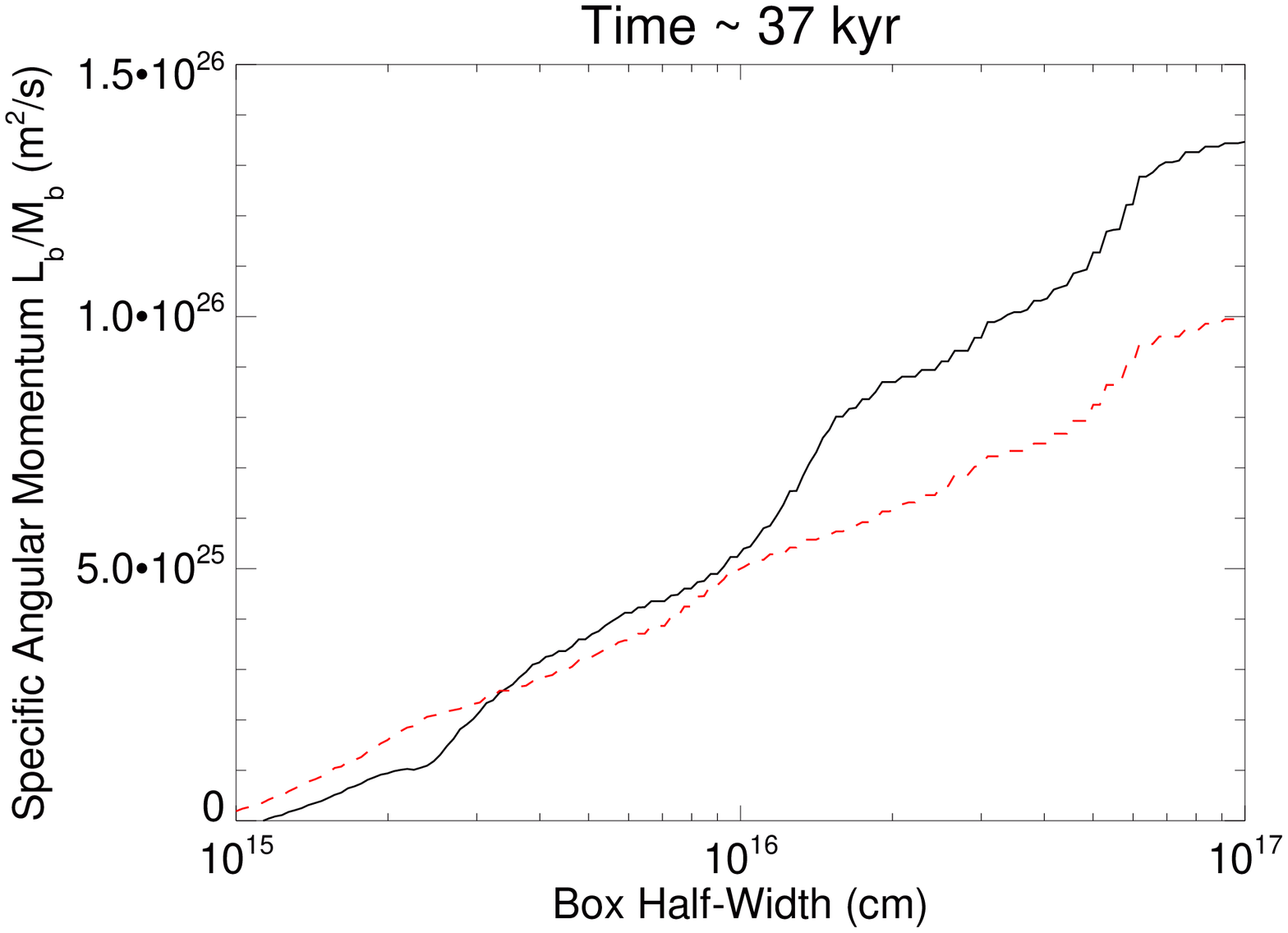}
\caption{The integrated gas specific angular momentum. The aligned case is shown in black solid curve while the orthogonal case is in red dashed.
Both are for $\lambda=4$ case, at a representative time $t\approx
37$~kyr.}
\label{spAMbox}
\end{figure}

Interestingly, the magnetic torque is not smaller at all radii when
the field is tilted $90^{\circ}$. At large radii ($\gtrsim 5\times
10^{16}$~cm), the orthogonal rotator, produces a stronger magnetic torque (see \ct{MouschoviasPaleologou1979}) than the one in the aligned rotator. The reverse is true as one moves to the inner region ranging from $\sim 1\times 10^{16}$~cm to $\sim 5\times 10^{16}$~cm (\ct{Mouschovias1991}). Moreover, the magnetic torque inside the innermost $\sim 8\times 10^{15}$~cm~$\approx 600$~AU region is comparable for both the aligned and orthogonal cases, which somewhat contradicts  the results of \citet{Joos+2012}. In other words, the magnetic braking is more efficient in outer regions of the core, but weaker close in, for the larger misalignment cases. Therefore, as the gas makes its way onto the binary stars when the field misalignment is large, the gas first loses more angular momentum relative to the aligned case, and then loses less closer in, because of both weaker magnetic braking and lack of magnetically-driven outflow. This is also consistent with the distribution of gas specific angular momentum shown in Fig.\ref{spAMbox}, where the orthogonal case is below the aligned case in the outer region, and yet the opposite is true in the inner region.
These subtle competing mechanisms explain the broad similarity in the binary orbital angular momentum we present in Fig.~\ref{lmd4am}. 

\subsection{Disk and Outflow}
\label{diskoutflow}

\begin{figure}
\epsscale{1.0}
\plottwo{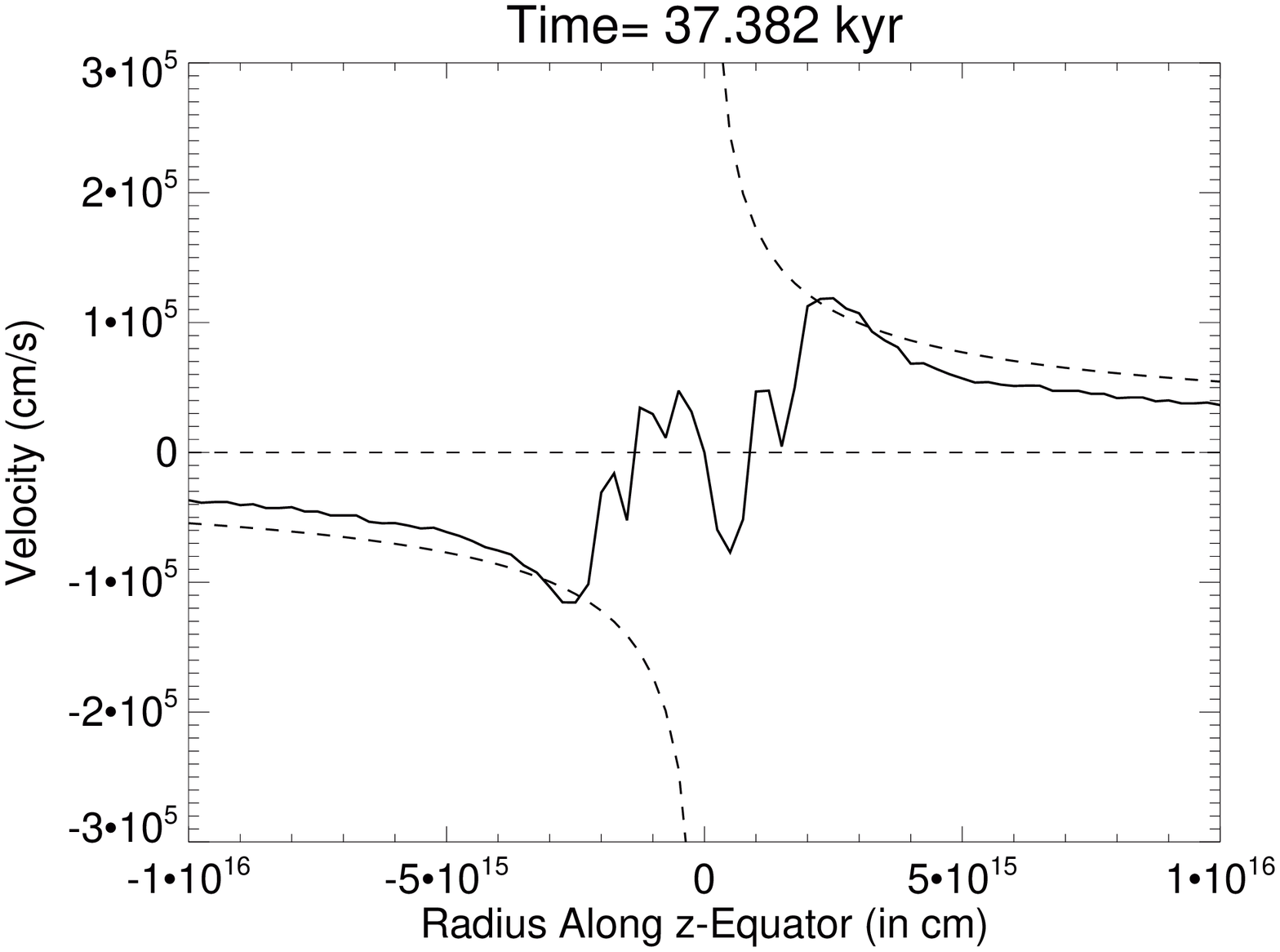}{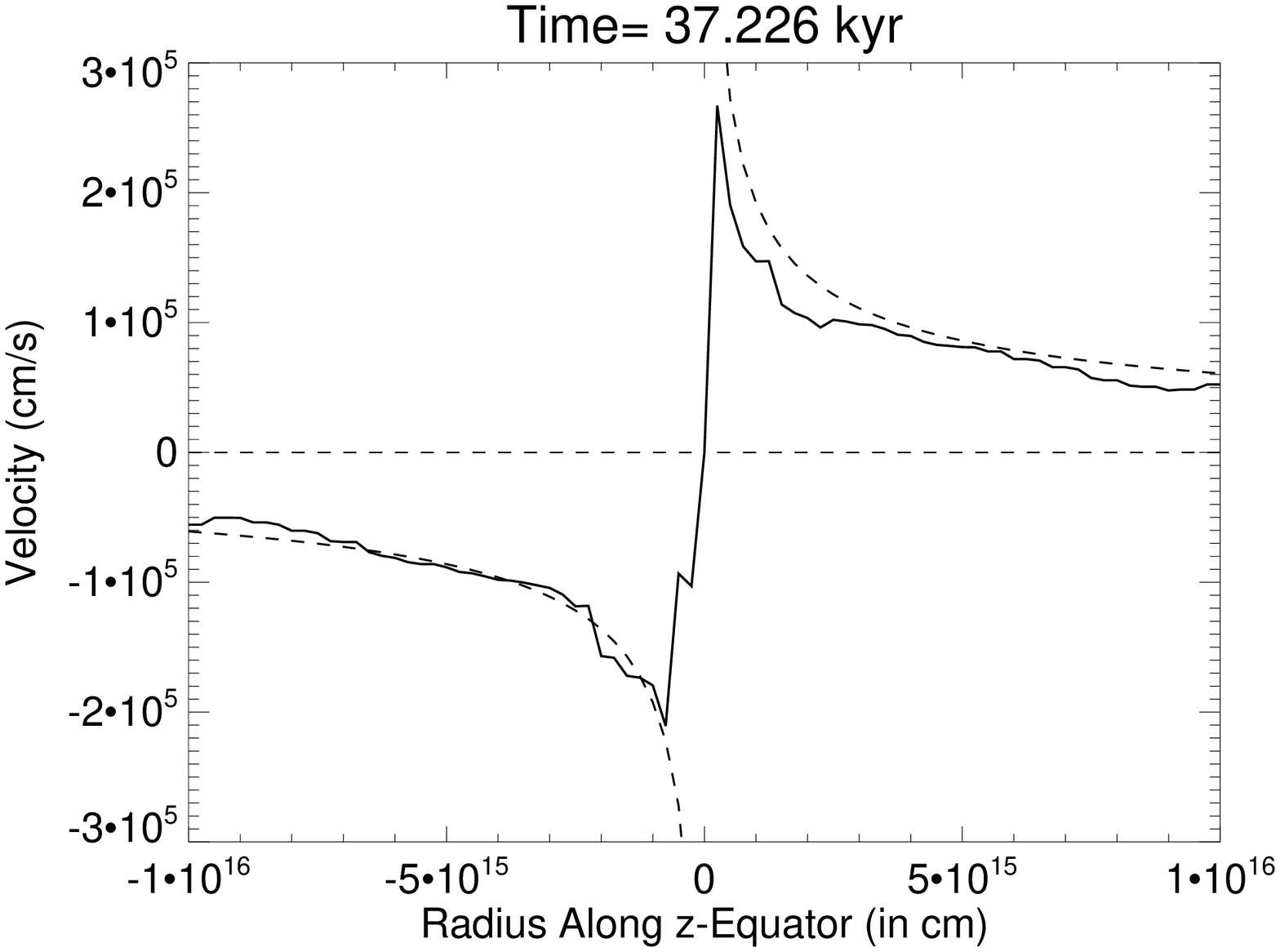}
\caption{The distribution of azimuthal velocity on the equator along the midline of two binary seeds for the $\lambda=4$ cases, at a representative time $t\approx 37$~kyr. Left panel is the aligned case and right being the orthogonal case. The estimated Keplerian profile is shown in dashed curve.}
\label{VelPro}
\end{figure}

The misalignment between the magnetic field and the rotation axis has
a noticeable effect on the disk morphology as well. In the aligned
cases, ZL13 showed that strong magnetic braking produces non-Keplerian
pseudo-disks in the circumbinary region (for $\lambda \lesssim 8$). In
contrast, we find near-Keplerian circumbinary disks in cases with
large field misalignment, even for mass-to-flux ratios as small as
$\lambda \sim 4$. Fig.~\ref{VelPro} plots the distribution of
azimuthal velocity $v_{\phi}$ on the equator along the midline of the
two seeds, where the orthogonal case shows a $\sim 400$~AU ($6\times
10^{15}$~cm) size disk that follows the estimated Keplerian
curve\footnote{Only the mass of the binary stars is used for
  calculating the Keplerian profile} in the circumbinary region. The
same region is occupied by sub-Keplerian structures in the aligned
case for the same level of magnetization ($\lambda=4$). Note that the
strength of magnetic torque does not differ much in both cases across
$10^2$~AU scale (see Fig.~\ref{figtorq}). However, the large DEMS
(Decouple-Enabled Magnetic Structures, about $\lesssim 400$~AU in size
at $t\sim37$~kyr) destroys any rotationally-supported structure, which
is an unavoidable magnetically-dominated feature when field is aligned
with the rotation axis (\ct{Zhao+2011}; \ct{Krasnopolsky+2012}).

\begin{figure}
\epsscale{1.0}
\plotone{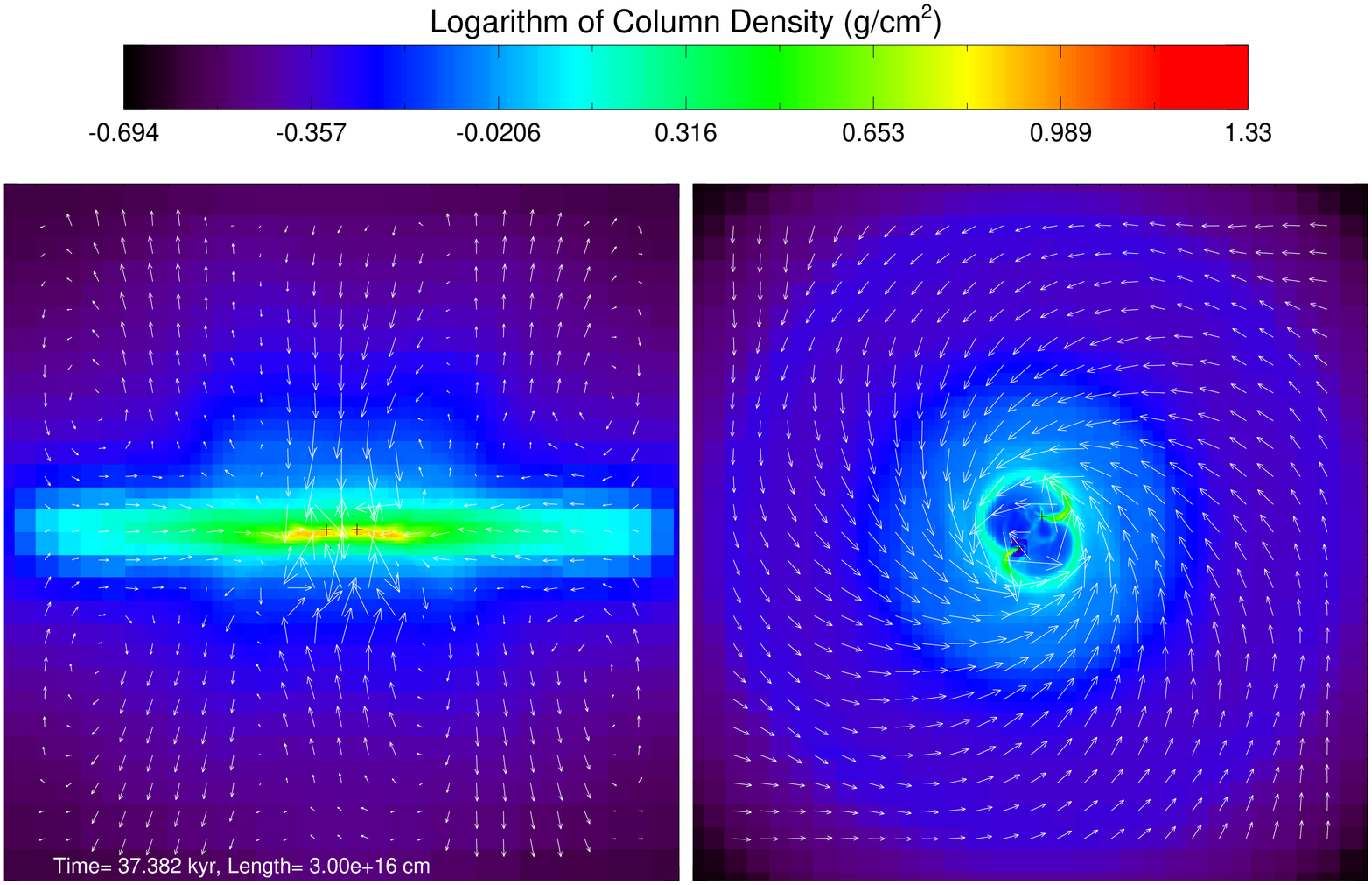}
\plotone{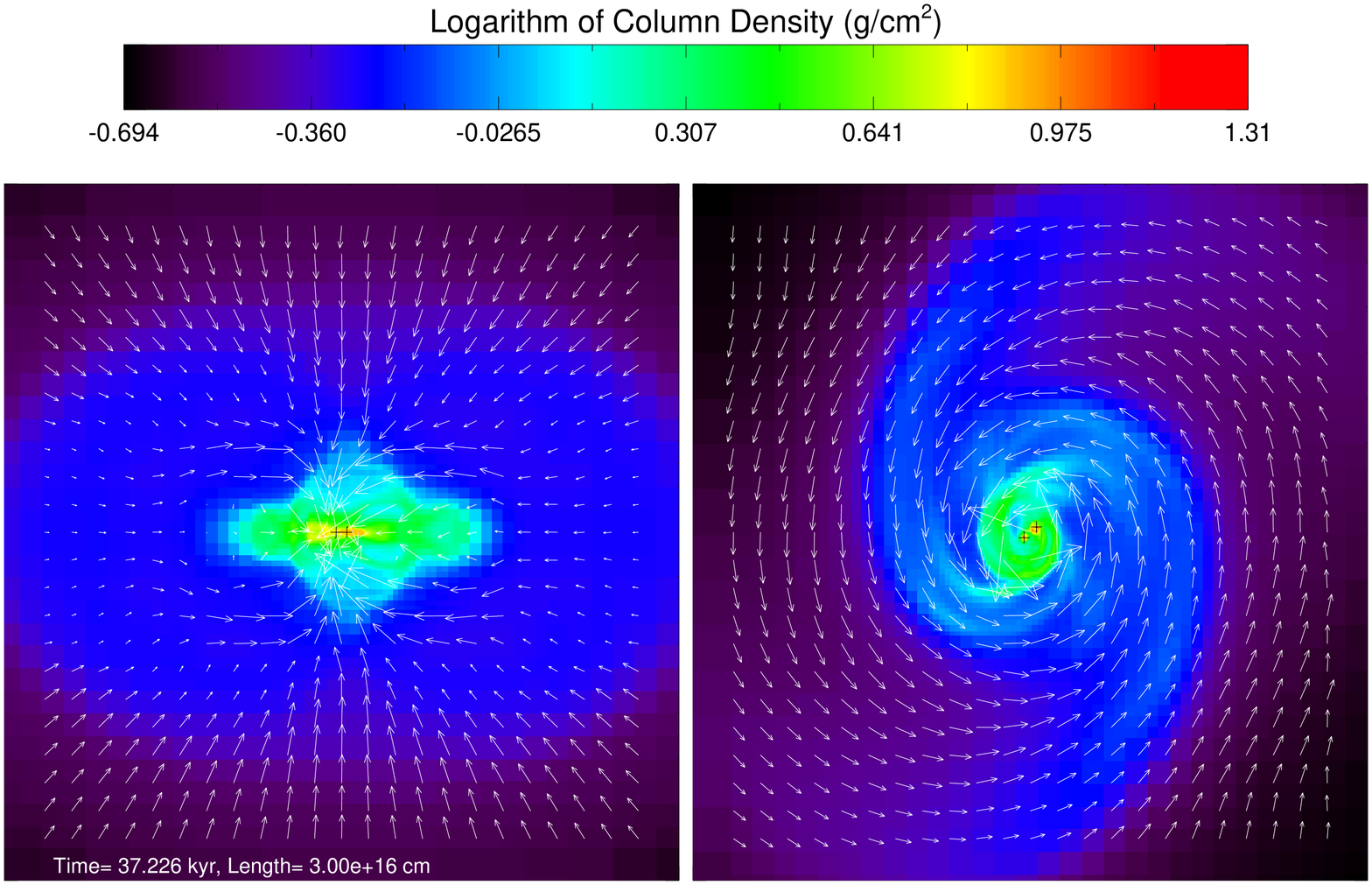}
\caption{The column density and velocity field in both edge-on (left
  panels) and face-on (right) view for the $\lambda=4$ cases, at a representative time $t\approx 37$~kyr. Upper panel is the aligned case and bottom being the orthogonal case. The length of region is $3\times 10^{16}$~cm.}
\label{columnplt}
\end{figure}

The DEMS also account for the survival of circumstellar disks. In ZL13, we find that circumstellar disks are suppressed even for relatively weak magnetic field ($\lambda\lesssim 16$), as soon as DEMS dominate the inner region close to the binary stars. As we tilt the magnetic field, the DEMS become less prominent. Hence the circumstellar disks start to survive even for $\lambda=8$ with tilt angle $\theta_0=45^{\circ}$, and for the early time of $\lambda=4$ with $\theta_0=90^{\circ}$ when the magnetic tension force is relatively weak. The effect of field misalignment on the DEMS is also obvious by comparing the right column plots in Fig.~\ref{columnplt}. Our study shows that large misalignment between the magnetic field and rotation axis can suppress the magnetic dominated structures (DEMS) to emerge in the central $\sim 10^2$~AU region - the cradle of infant protostellar and protobinary disks.


Fig.~\ref{columnplt} (left column) also shows the effect of field
misalignment on outflow structures, which is consistent with our
discussion so far and Fig.~\ref{figtorq}. We do not observe any
obvious outflow structure in the orthogonal cases. The aligned case
has large magnetically-driven outflow launching regions both above and
below the equator (see also ZL13). The same regions are dominated by
gas infall for the orthogonal case. Interestingly, there are two
prominent spirals in the face-on view (lower-right panel); in 3D, 
they are 
the snail-shell like structures discussed in \citet{Li+2013}. 

\subsection{Mass Ratio}
\label{massratio}

\begin{figure}
\epsscale{0.9}
\plotone{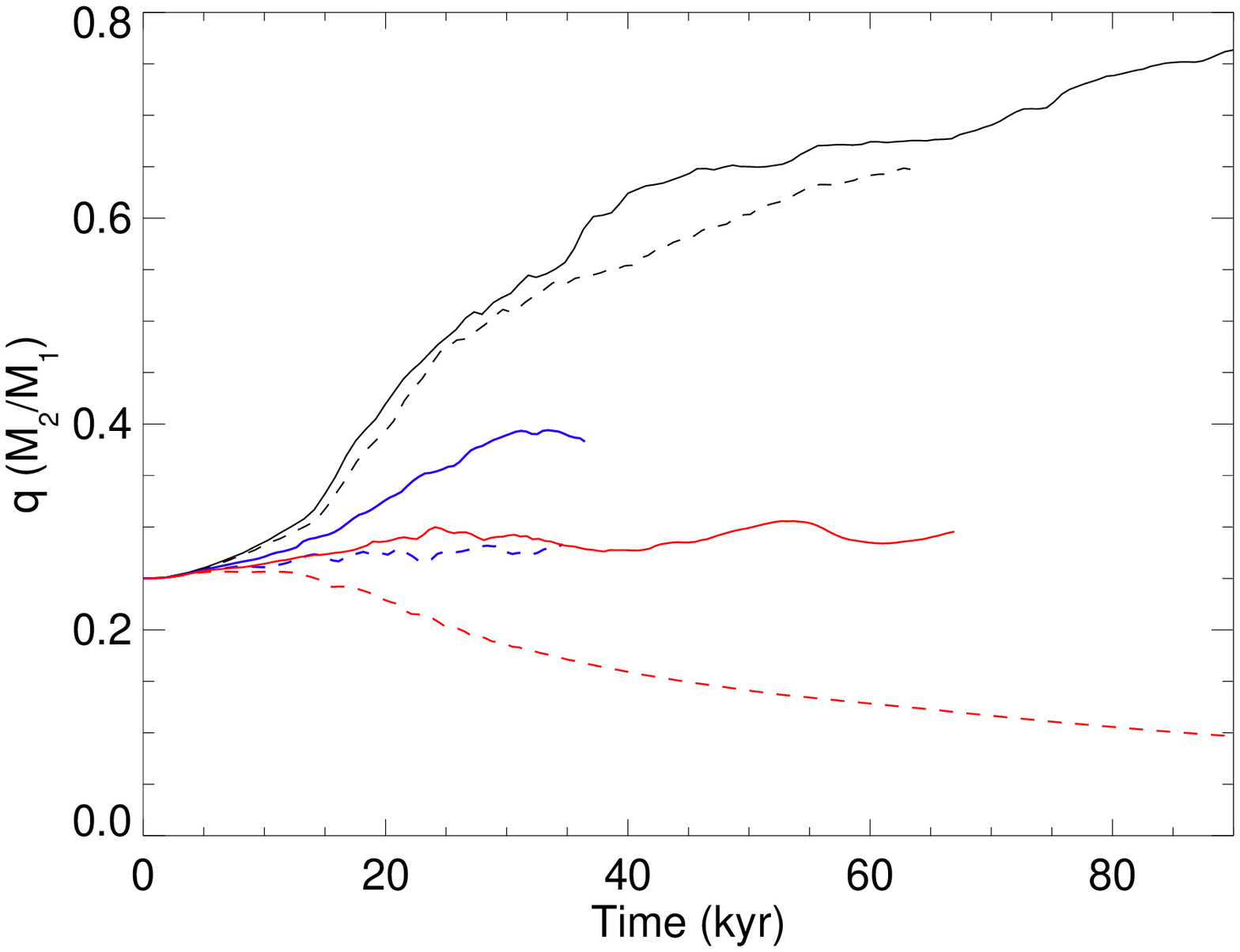}
\caption{Evolution of mass ratio for an initially unequal mass binary
  of $q=0.25$. Aligned cases are shown in solid lines and orthogonal 
  cases are in dashed curves. The black, blue and red lines are for $\lambda=32$, $4$, and $2$, respectively.}
\label{masrat}
\end{figure}

Besides orbital separation, another fundamental quantity that
characterizes a binary system is the mass ratio $q=M_2/M_1$ 
($M_1$ and $M_2$ are the mass of the primary and secondary, 
respectively).  We have previously shown that the mass ratio 
is strongly affected by magnetic braking in the aligned case 
(ZL13). In particular, we found that the magnetic braking 
can slow down or even eliminate the increase of $q$ 
(e.g. growth towards equal mass) with time for initially unequal mass binaries by weakening or 
suppressing the preferential mass accretion onto the secondary. 
We have carried out calculations similar to ZL13 for unequal 
binaries of initial $q=0.25$ but with both $\theta_0=0^\circ$ 
and $90^\circ$. Fig.~\ref{masrat} compares the time evolution 
of mass ratio $q$ for both the aligned and orthogonal cases 
for different values of mass-to-flux ratio $\lambda$. It is clear 
that tilting the magnetic field not only suppresses the accretion onto the secondary compared to the hydrodynamic case, but can also increase the accretion onto the primary compared to the aligned case.
In the strongest field case of $\lambda=2$, the mass ratio 
actually decreases with time, the opposite of what one may 
expect based on hydrodynamical simulations (\citealt{BateBonnell1997,Kratter+2010}; ZL13). It therefore appears that magnetic misalignment makes it easier for highly uneven mass binaries to preserve their initial mass 
ratios during the protostellar mass accretion phase. Why this 
is the case is unclear; it deserves further investigation, which 
is beyond the scope of this paper.

\section{Summary and Discussion}
\label{discuss}

In this paper, we have investigated numerically the effect of misalignment between the magnetic field and rotation axis on protobinary evolution during the mass accretion phase, in direct comparison to our previous calculations with aligned magnetic field (\ct{ZhaoLi2013}). We have found that a misaligned magnetic field is more efficient  than the aligned field at tightening the binary separation. The fundamental reason is that  field misalignment suppresses the magnetically-driven outflows. This suppression has three consequences. First, the weaker gas outflow 
allows more of the infalling gas to accrete onto the binary stars, which increases the binary mass relative to the aligned field case. Secondly, the weaker magnetically-driven outflow carries away less angular momentum from the infalling gas in the midplane. This second effect on its own would lead to wider binary orbits. However, there exists a competing mechanism where the same parcel of gas at larger radii has experienced stronger magnetic braking when field misalignment is greater. Thus the excess accreted gas provides only slightly bigger total angular momentum to the binary orbit. Thus  the magnetic field misalignment allows faster mass accretion  accompanied by a much smaller increase in angular momentum.
Thus the binary separations in all misaligned runs are smaller than their counterparts in \citet{ZhaoLi2013}. 

The efficient tightening of the binary orbit by magnetic field misalignment lowers the field strength required for the same degree of migration. As mentioned in the introduction, the distribution of binary orbital separations during the earliest Class 0 phase of star formation may differ from those at later times. In particular, there may exist a gap free of Class 0 binaries with separations between $150-550$~AU \citep{Maury+2010, Enoch+2011}, which is not present in the Class I or later phases\footnote{A recent SMA survey by \citet{Chen+2013} finds more evidence for Class 0 binarity, but a few of the apparent multiple systems may not evolve into binaries based on their sensitivity and resolution.}. We previously showed that magnetic braking is an efficient way of migrating protobinaries born on wider  orbits into this apparent separation gap (\ct{ZhaoLi2013}).
The results presented here indicate that to reach the same level of migration as in the aligned field cases, one may only require a magnetic field that is half as strong if it is misalignment with the rotation axis. 

Our proposed mechanism of binary migration due to a tilted magnetic field may be favorable in the context of disk formation in magnetized prestellar cores. Recent numerical work has shown that rotationally supported disks may form 
under the combination of low magnetic field strengths and large field misalignment \citep{Li+2013, Joos+2012, Krumholz+2013}. The CARMA sample of \citet{Hull+2013} shows that the distribution of the angle between the magnetic field and jet/rotation axis is consistent with being random. This would indicate that in $50\%$ of the sources the two axes are misaligned by a large angle ($\gtrsim 60^{\circ}$). If true, the large misalignment would allow disk formation in moderately ($\lambda \gtrsim 4$) magnetized prestellar cores (\ct{Li+2013}). Therefore for cores that form binary systems, the misaligned and weaker magnetic field would enable  efficient migration to fill the gap observed in \citet{Maury+2010}, while still allowing for the formation of $10^2$-AU scale disks
in the Class 0 phase. Furthermore, the survival of rotationally-supported circumbinary disks through tilting the magnetic field could explain some recent SMA and ALMA binary observations. \citet{Takakuwa+2012} identify a circumbinary disk with possible spiral structures around protobinary system L1551 NE, which may arise from local perturbations due to magnetic field misalignment or supersonic turbulence. We will discuss the effects of turbulence in a future study.

\acknowledgments
We thank P. Arras, and A. Maury for useful discussions and P. Wang for advice on the ENZO code. This work is supported in part by NASA NNX10AH30G. Support for this work was also provided by NASA through Hubble Fellowship grant \#HF-51306.01 awarded by the Space Telescope Science Institute, which is operated by the Association of Universities for Research in Astronomy, Inc., for NASA, under contract NAS 5-26555.

\end{document}